\documentclass[12pt]{article}

\usepackage{newtxtext,newtxmath}

\usepackage{graphicx}
\graphicspath{{Nature/}{Nature/Figures/}{Nature/suppl_figures/}}

\usepackage[letterpaper,margin=1in]{geometry}

\usepackage{setspace}
\date{}

\usepackage[super,sort&compress]{natbib}

\usepackage{url}

\usepackage{authblk}

\usepackage{xspace}
\newcommand{\sto}{SrTiO\textsubscript{3}\xspace}

\title{\textbf{Evidence for a textured polar state in strained SrTiO$_3$}}

\author[1,2]{Huaiyu~Hugo~Wang}
\author[1,2]{Ernesto~Flores}
\author[1,2]{Jade~Stanton}
\author[1,2]{Gal~Orenstein}
\author[3]{Peter~R.~Miedaner}
\author[4]{Laura~Foglia}
\author[1]{Maya~Martinez}
\author[1,2]{David~A.~Reis}
\author[5]{Roman~Mankowsky}
\author[5]{Mathias~Sander}
\author[5]{Henrik~Lemke}
\author[5]{Shih-Wen~Huang}
\author[5]{Serhane~Zerdane}
\author[3]{Keith~A.~Nelson}
\author[1,2,*]{Mariano~Trigo}

\affil[1]{Stanford Institute for Materials and Energy Sciences, SLAC National Accelerator Laboratory, Menlo Park, CA 94025, USA}
\affil[2]{Stanford PULSE Institute, SLAC National Accelerator Laboratory, Menlo Park, CA 94025, USA}
\affil[3]{Department of Chemistry, Massachusetts Institute of Technology, Cambridge, MA 02138, USA}
\affil[4]{Elettra-Sincrotrone, Trieste 34149, Italy}
\affil[5]{Paul Scherrer Institut, 5232 Villigen PSI, Switzerland}
\affil[*]{Corresponding author. Email: mtrigo@slac.stanford.edu}

\begin{document} 

\maketitle

\begin{abstract}
Quantum materials can harbour hidden phases whose microscopic structures differ from conventional ordered states while reproducing their macroscopic signatures, making them easy to miss. Strontium titanate is a longstanding puzzle of this kind: on cooling it shows every hallmark of an incipient ferroelectric, yet never orders, and is usually described as a quantum paraelectric in which fluctuations suppress ferroelectricity. Here we combine uniaxial strain, single-cycle terahertz excitation and femtosecond x-ray scattering to measure the polar collective modes of strontium titanate as a function of momentum and strain. Under modest tensile strain, a new infrared-active mode appears at finite wavevector, while the zone-centre transverse optic mode does not soften. The strain-induced polar regime is therefore not reached through the zone-centre instability of a uniform ferroelectric, and is consistent with a polarization textured on nanometre length scales, either static or fluctuating. These results suggest that the polar instability in quantum paraelectric strontium titanate may lie at finite wavevector, and show how finite-momentum collective excitations can reveal hidden phases in quantum materials.
\end{abstract}


Identifying new phases of matter is a foundational goal of condensed matter physics, but phases that break symmetry only locally or at finite length scales can be nearly invisible to the macroscopic probes --- susceptibility, transport, diffraction --- that have historically defined the phase diagrams. Such hidden phases can reproduce the thermodynamic fingerprints of conventional ordered states while differing from them in their microscopic structures~\cite{Mydosh2011,Stojchevska2014,Basov2017,Kogar2020,Disa2021}, and distinguishing the two requires probes sensitive to collective excitations away from the Brillouin zone centre~\cite{schulke2007electron,Ament2011,Abbamonte2025}. Strontium titanate (SrTiO$_3$) is a striking case in point. On cooling, its dielectric susceptibility diverges~\cite{Weaver1959} and its zone-centre polar mode softens~\cite{Yamada1969} --- the textbook signatures of an incipient ferroelectric --- yet long-range ferroelectric order never sets in, even at the lowest temperatures. This behavior has for decades been attributed to quantum fluctuations that suppress an otherwise conventional ferroelectric instability, making SrTiO$_3$ the archetypal quantum paraelectric~\cite{Muller1979,Rowley2014}. An alternative possibility, however, is that the instability is not toward homogeneous ferroelectricity, but toward a spatially modulated polar state whose order parameter develops at finite wavevector and is therefore hidden from the macroscopic probes that shaped the canonical picture~\cite{Muller1991,Zubko2007,Guzman-Verri2023,Orenstein2025}. Here we use externally applied uniaxial strain, single-cycle terahertz excitation, and femtosecond x-ray scattering to test these scenarios by measuring the polar collective modes of SrTiO$_3$ as a function of momentum under strain. 
We find that moderate tensile strain, which earlier studies showed drives \sto into a polar regime \cite{Burke1971,Uwe1976,Li2025}, produces a new infrared-active mode at finite wavevector, while the zone-centre transverse optical mode does not soften. This is inconsistent with a homogeneous ferroelectric transition at the strains studied and points to a polarization textured on nanometre length scales, either static or fluctuating. If this regime is continuously connected to the unstrained state, quantum paraelectric \sto may be better viewed as the disordered side of a transition into a textured polar state than as a ferroelectric suppressed by quantum fluctuations.

\begin{figure} 
	\centering
	\includegraphics[width=1\textwidth]{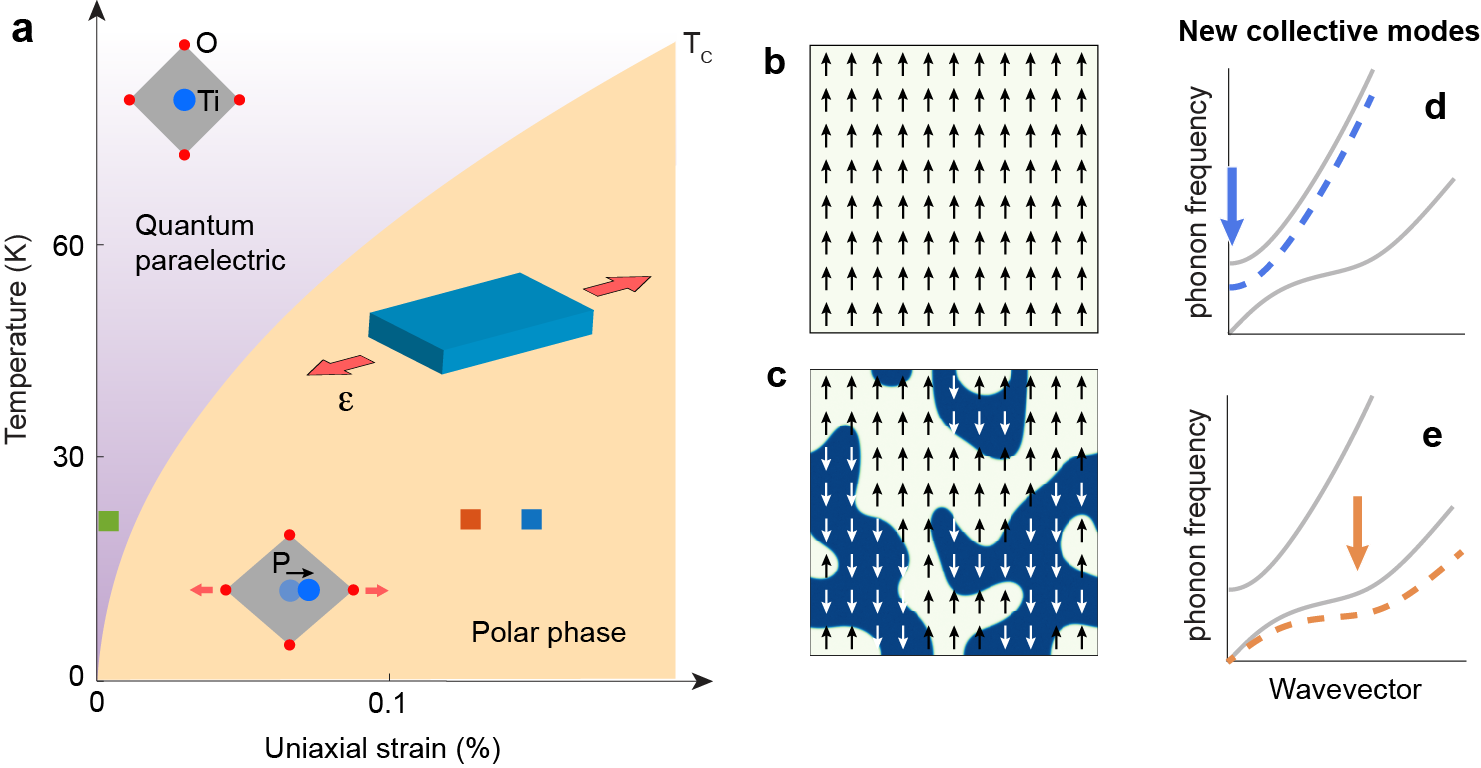} 
	\caption{\textbf{Quantum phases in strained SrTiO$_3$.} (\textbf{a}) Phase diagram of SrTiO$_3$. The unstrained quantum paraelectric phase is shown in purple. The orange shaded region represents a polar phase conventionally identified as a ferroelectric phase. $T_{c}$ is the critical temperature. 
    Green, red, and blue squares mark the three strain conditions, $0,(1.30\pm0.01)\times10^{-3}$ and $(1.45\pm0.01)\times10^{-3}$, respectively measured in this work (Figs. \ref{fig2}, \ref{fig3}, \ref{fig4} and \ref{fig5}) at $T = 20$~K, plotted against the uniaxial strain $\epsilon$ assuming $\sigma = 0.2$ (Methods). The phase boundary is the measured uniaxial-strain $T_c(\epsilon)$ \cite{Li2025}.
    (\textbf{b}) Illustration of a conventional ferroelectric with a homogeneous polarization (black arrows) and globally broken inversion symmetry.
    (\textbf{c}) Illustration of a textured polar phase where the polarization is spatially inhomogeneous (black and white arrows) with locally broken inversion symmetry.
    (\textbf{d}) - (\textbf{e}) illustrations of the behavior of the phonon branches in the two phases. Upper and lower gray lines indicate the optical and acoustic modes in the paraelectric phase. Dashed blue and orange lines represent the renormalized mode that appears in the new regime, (\textbf{b}) and (\textbf{c}), respectively.
    }
	\label{fig1} 
\end{figure}

Although \sto is traditionally characterized as a quantum paraelectric in which quantum fluctuations prevent ferroelectric order\cite{Muller1979}, this characterization rests largely on macroscopic susceptibility measurements, which cannot distinguish a uniform paraelectric ground state from one with local symmetry breaking at finite wavevector. The material exhibits quantum criticality\cite{Rowley2014,Coak2020} and extreme sensitivity to non-thermal perturbations including ultrafast pulses\cite{Li2019,Nova2019}, doping\cite{Bednorz1984,Fauque2026}, oxygen substitution\cite{Itoh1999}, and strain\cite{Haeni2004,Li2025,Uwe1976}, suggesting proximity to other broken-symmetry phases. This extreme tunability raises a question central to any full picture of quantum paraelectricity: which broken-symmetry phase is the system closest to, and is it already partially realized at low temperature?
Fig.~\ref{fig1}a shows a schematic phase diagram\cite{Coak2020,Guzman-Verri2023} of \sto under uniaxial strain; analogous diagrams can be constructed under hydrostatic pressure, epitaxial strain, or chemical substitution\cite{Rowley2014,Coak2020,Guzman-Verri2023}. The purple region is the quantum paraelectric phase, where the susceptibility increases on cooling but the system is conventionally assumed to remain inversion-symmetric, represented schematically by the centrosymmetric Ti-O square in the inset. The orange region represents the polar phase reached under epitaxial\cite{Haeni2004,Schlom2007} or uniaxial\cite{Uwe1976,Burke1971,Fujii1987,Li2025,Xu2020} strain, conventionally identified with a ferroelectric transition. The phase boundary in Fig. \ref{fig1}a is taken from freestanding membranes \cite{Li2025}, whereas our sample is a bulk crystal and thus the boundary serves as a guide rather than a precise reference for our sample, whose placement in the polar regime is supported by earlier measurements on bulk crystals \cite{Uwe1976,Burke1971,Fujii1987}.

The canonical interpretation of this phase diagram is the displacive picture of Fig.~\ref{fig1}b,d: a homogeneous spontaneous polarization, represented in \ref{fig1}b by the black arrows, develops below $T_c$, accompanied by a soft ferroelectric mode at small wavevector and a strong renormalization of the zone-centre phonon dispersion\cite{Cochran1960}, highlighted by the blue arrow in Fig. \ref{fig1}d. Under this picture, the orange region is a uniform ferroelectric --- its signature is a new collective mode shown by the dashed blue line in \ref{fig1}d --- and the purple region is a uniform paraelectric held back from ordering only by quantum fluctuations. 
In an alternative scenario\cite{Muller1991,Guzman-Verri2023}, shown schematically in Fig.~\ref{fig1}c,e, the symmetry breaks locally but not globally, producing a spatially inhomogeneous polarization (opposite pointing arrows in Fig.~\ref{fig1}c) and a new phonon branch at \emph{finite} wavevector (dashed orange line in Fig.~\ref{fig1}e), rather than a new soft mode at the zone centre. Both scenarios can be compatible with a large macroscopic susceptibility as the relevant mode softens, so dielectric measurements alone cannot distinguish them. 
A direct test requires measuring the polar phonon spectrum at finite wavevector in the strained regime, which we report below.

\begin{figure} 
	\centering
	\includegraphics[width=0.65\textwidth]{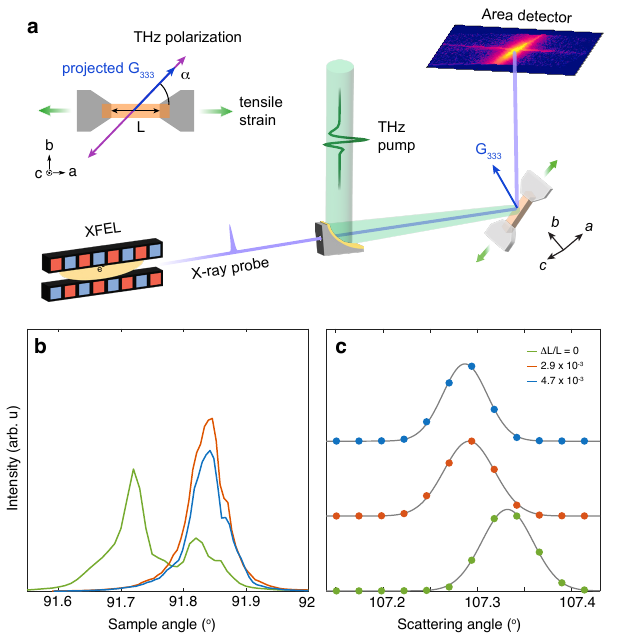} 
	\caption{\textbf{Structural evolution as a function of tensile strain} (\textbf{a}) 
    Schematic illustration of the experimental setup at the Bernina station of the Swiss FEL. A THz pulse with a spectrum centered around 0.5 THz (Fig.~\ref{fig:s1}) is focused onto a 50~$\mu$m-thick SrTiO$_3$ sample mounted on a Razorbill CS130 strain cell. Snapshots of the lattice are measured by diffraction of a 50 fs x-ray pulse with photon energy of 10 keV. The sample temperature was 20 K. Inset: Illustration of THz pump polarization and uniaxial strain directions with respect to crystallographic axes of the sample. 
    (\textbf{b})
    Rocking curve of the (3,3,3) Bragg peak for representative tensile strains. The change in the effective sample length from the nominal unstrained condition is labeled in the inset of (\textbf{a}). The curves are labeled by the measured displacement of the strain-cell relative to that of the green line (further details in Methods). 
    (\textbf{c}) The diffraction peak profile on the detector as a function of scattering angle. The solid curves are gaussian fits.}
	\label{fig2}
\end{figure}

We applied uniaxial strain at cryogenic temperatures to tune quantum-paraelectric \sto toward its polar phase. To characterize this phase, we probed the mesoscopic polar excitations in the system using ultrafast x-ray scattering at the Bernina instrument of the SwissFEL facility\cite{Prat2020}. A single-cycle terahertz (THz) pump pulse resonantly excited the soft polar modes of \sto, followed by a femtosecond hard x-ray diffraction probe, as shown in Fig.~\ref{fig2}a (Methods). Because the THz field couples to the lattice at first order, it excites primarily infrared-active phonons. The THz excitation is therefore selective to collective \emph{polar} modes, enabling direct mapping of their energy dispersion with exceptional wavevector and energy resolution\cite{Orenstein2025}. 
When combined with strain tuning as we show here, the collective modes constrain the microscopic nature of the transition.

Bulk single-crystal \sto with dimensions $5 \times 0.2\times 0.05$ mm$^3$ was mounted on a tunable strain-cell and cryogenically cooled to $T = 20$~K. Fig.~\ref{fig2}b shows rocking curves around the $(3,3,3)$ Bragg reflection (pseudo-cubic notation) taken for three strain-cell displacements $\Delta L$. Here $\Delta L$ is the nominal change in the sample length obtained through the change in capacitance of the strain device. The strain that transfers to the sample surface illuminated by the x-rays is smaller (see Methods). The green, red, and blue lines correspond to increasing tensile strain along the cubic $[100]$ direction. The green curve exhibits two peaks originating from distinct tetragonal domains that typically develop in unstrained \sto below the $T=105K$ antiferrodistortive phase transition\cite{lytle1964x,haywardCubictetragonalPhaseTransition1999,Shirane1969,Orenstein2025}. As $\Delta L$ increases, the peak at $91.72^\circ$ is suppressed (Fig.~\ref{fig2}b), indicating detwinning of the sample under tensile strain towards a monodomain configuration where the tetragonal axis is expected along the tensile direction\cite{chang1970direct,MULLER1970monodomain}. 

With increasing tensile strain, the peak at $91.82^\circ$ shifts to larger angles, reflecting the modification of the $(3,3,3)$ interplanar spacing, $d_{(3,3,3)}$. Fig.~\ref{fig2}c shows the x-ray scattering intensity as a function of the scattering angle for each strain, with the crystal orientation fixed to the maximum of the rocking curve corresponding to the same domain (the green curve was measured at $91.82^\circ$). The Bragg reflection shifts to lower scattering angles with increasing $\Delta L$, confirming an increasing $d_{(3,3,3)}$ under tensile strain. From these shifts, we obtain the changes in interplanar spacing for the red and blue curves referenced to the green curve, $\Delta d_{(3,3,3)}/\bar{d}_{(3,3,3)}=2.60 \times 10^{-4}$ and $2.89\times 10^{-4}$, respectively, where $\bar{d}_{(3,3,3)}$ is the nominal value at this temperature. Complementary calibration of the applied strain based on capacitance measurements of the strain device is described in the Strain details subsection of Methods.

\begin{figure}
	\centering
	\includegraphics[width=1\textwidth]{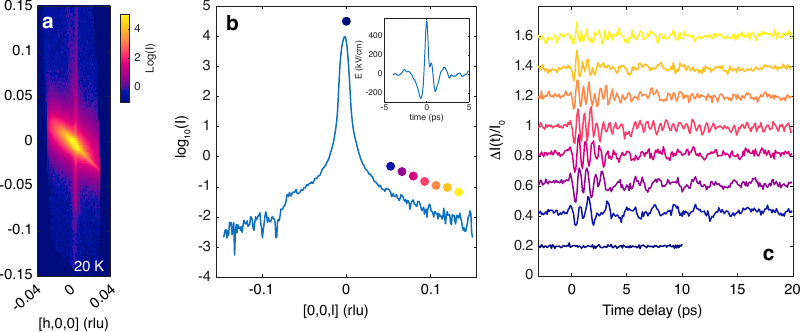} 
	\caption{\textbf{Coherent polar response under tensile strain.} (\textbf{a}) X-ray scattering intensity in reciprocal space around the (3,3,3) Bragg peak in the [$h$,0,$l$] plane at 20K. (\textbf{b}) X-ray intensity along [0,0, $l$] direction near the (3,3,3) Bragg peak (logarithmic scale). Inset: electro-optic sampling trace of the THz pulse. (\textbf{c}) Relative x-ray intensity $\Delta I(t)/I_0$ for strain at $\Delta L/L = 4.7 \times 10^{-3}$, as a function of the delay between the THz and x-ray pulses, where $\Delta I(t)=I(t)-I_0$ and $I_0$ is the x-ray intensity before the THz excitation. The vertically offset lines correspond to different representative wavevectors along the $[0,0, l]$ direction indicated by the matching color dots in (\textbf{b}).}
	\label{fig3} 
\end{figure}

\begin{figure} 
	\centering
	\includegraphics[width=0.8\textwidth]{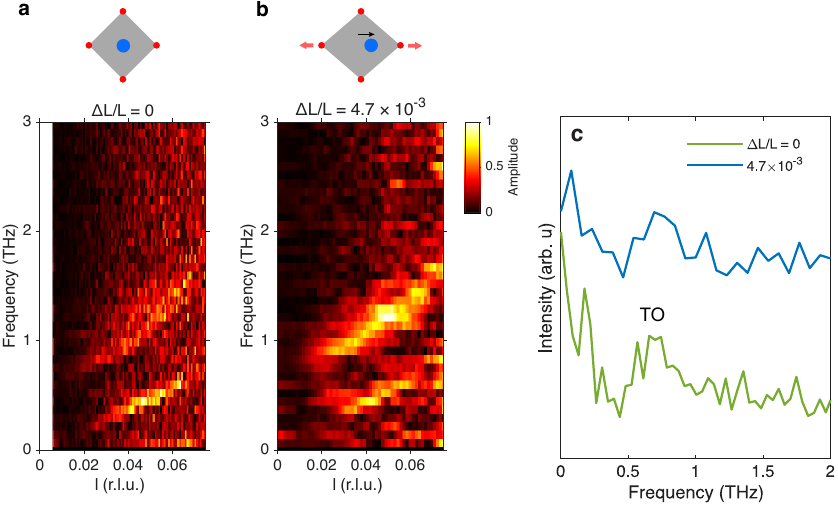} 
	\caption{\textbf{Near zone centre polar phonon dispersion} (\textbf{a}-\textbf{b}) The color maps represent the magnitude of the Fourier transform (FT) of $\Delta I(t)/I_0$ for the $\Delta L/L = 0$  (a) and $\Delta L/L = 4.7 \times 10^{-3}$ (b) at $20$~K, computed from time-domain data like those in Fig.~\ref{fig3}c. The region $l < 0.01$ rlu is affected by detector saturation at the Bragg peak (Methods). (\textbf{c}) The magnitude of the FT at $l=0.01$ rlu for the two strain configurations in (\textbf{a}-\textbf{b}). Curves are displaced vertically for clarity.
    }
	\label{fig4} 
\end{figure}

To access the polar collective modes, we combine ultrafast x-ray scattering with single-cycle THz excitation. We stress that the pump pulse does not produce a structural phase transition, instead it generates coherent oscillations of the polar modes in the linear response regime\cite{Orenstein2025}, and the collective modes observed are those near equilibrium.
Fig.~\ref{fig3}a shows a reciprocal-space map of the x-ray intensity around the $(3,3,3)$ Bragg reflection at $20$~K (pseudo-cubic notation). In this basis, the sample surface normal is along the $l$ direction, which induces an elongated streak from the crystal surface truncation that is independent of strain and temperature (see \ref{fig:s3} and \ref{fig:s4}). The associated logarithmic intensity plot along $[0,0,l]$ is shown in Fig.~\ref{fig3}b. 
Fig.~\ref{fig3}c shows representative traces of the time evolution of the relative x-ray intensity following THz excitation, which appears predominantly along the [$0,0,l$] streak. This feature suggests that the detection of the oscillatory signal is enhanced by heterodyne mixing between the phonon and the crystal truncation streak\cite{Shayduk2022}. The same termination of the lattice at the surface relaxes crystal-momentum conservation along $l$, which may also allow the long-wavelength THz field to drive polar modes at finite wavevector along this direction. 
The THz pulse, measured by electro-optic sampling as described in Methods, is shown in the inset of Fig.~\ref{fig3}b. The data in Fig.~\ref{fig3}c were taken at the largest applied strain (blue traces in Fig.~\ref{fig2}), with individual traces corresponding to representative wavevectors indicated in Fig.~\ref{fig3}b; measurements at other strains are provided in \ref{fig:s2}.
The oscillatory signal clearly varies significantly with wavevector. 
The response is strongest for $l \approx 0.06–0.13$ rlu, the range where the transverse optical (TO) and transverse acoustic (TA) branches are most strongly coupled\cite{Orenstein2025,Fauque2022}, lending the acoustic branch the polar character required for direct THz excitation. 
Notably, we do not observe any appreciable dynamics at the Bragg peak for any of the strains measured (dark blue line in Fig. \ref{fig3}c).
To test for zone-centre softening of the TO branch directly, we measured the response near the zone centre with high wavevector resolution. The color maps in Figs.~\ref{fig4}a-b represent the magnitude of the Fourier transform of the oscillatory signal at wavevectors along [$0,0,l$] near the zone-center.  
Both the TA and TO branches are excited down to $l = 0.01$ rlu, and the TO mode is essentially unchanged at low wavevector between the unstrained and most strained conditions (Fig. \ref{fig4}c). Together with the absence of a Bragg-peak response (Fig. \ref{fig3}c), this indicates no homogeneous polarization associated with the long-wavelength TO mode at the strains measured.

\begin{figure} 
	\centering
	\includegraphics[width=1\textwidth]{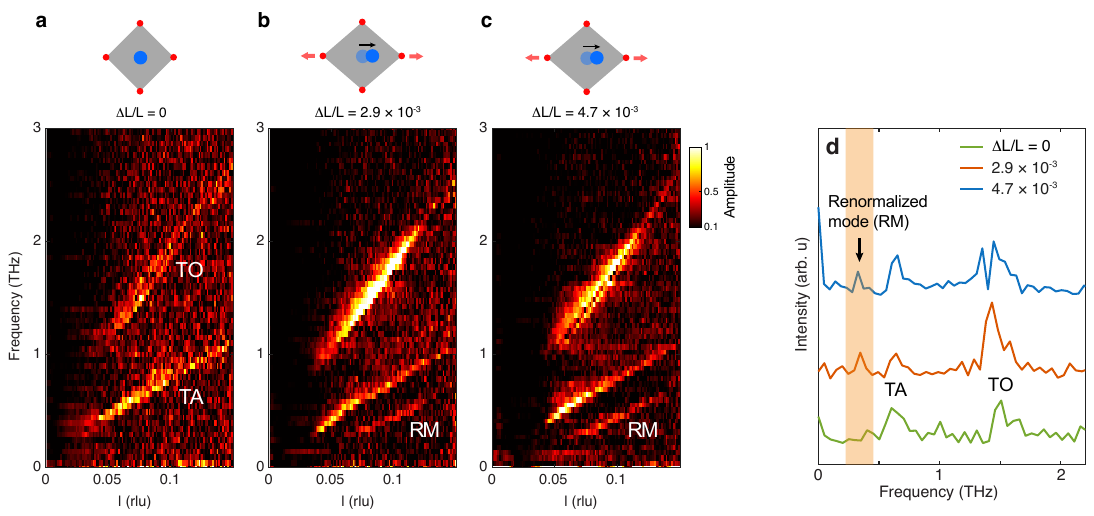} 
	\caption{\textbf{Collective polar modes of the textured state under strain.} (\textbf{a}-\textbf{c}) The color maps represent the magnitude of the Fourier transform of $\Delta I(t)/I_0$ for increasing tensile strain at $20$~K, computed from time-domain data like those in Fig.~\ref{fig3}c. The material is detuned from the Bragg condition. (\textbf{d}) The magnitude of the FT at $l=0.075$ rlu for the three strain configurations in (\textbf{a}-\textbf{c}). Curves are displaced vertically for clarity. The renormalized mode (RM) of vibration is labeled in (\textbf{b}, \textbf{d}).
    }
	\label{fig5} 
\end{figure}

Away from the zone centre, by contrast, the polar modes change markedly with strain.
The color maps in Figs.~\ref{fig5}a-c show similar plots as Figs. \ref{fig4}a-b reaching higher wavevectors along [$0,0,l$], for the three increasing strains presented in Fig. \ref{fig2}. 
These data, recorded slightly detuned from the Bragg condition, are optimized for larger wavevectors, $l \gtrsim 0.04$~rlu, while Fig.~\ref{fig4} covers the region near the zone centre (Methods).
At the lowest strain (Fig.~\ref{fig5}a), we observe two dispersive modes spanning $0.3-1$ THz and $1.2-2.5$ THz, associated with the TA and TO phonon branches, respectively; the THz field is polarized perpendicular to the wavevector (Methods), so only transverse modes are driven, and the two branches are distinguished by the linear approach of the lower one to zero frequency at $l \to 0$. With increasing tensile strain (Figs.~\ref{fig5}b and~\ref{fig5}c), a third mode, labeled renormalized mode (RM), appears below the TA branch (see also Extended Data 5). Note that the TA is the lowest frequency branch along $[0,0,l]$ in the unstrained system below the antiferrodistortive transition\cite{Courtens1993}. This holds for each tetragonal domain of the unstrained sample, which show only the TA peak at l=0.06 rlu (Fig.~\ref{fig:s6}); the new mode is thus not a domain-averaged branch revealed by detwinning. The new mode is resolved for $l \gtrsim 0.06$ rlu, whereas the TO mode is resolved down to $l = 0.01$ rlu (Fig. \ref{fig4}).
This change is apparent in the spectra for $l=0.075$~rlu shown in Fig.~\ref{fig5}d, where the green, red, and blue traces correspond to the lowest, intermediate, and highest strains, respectively. Compared with the lowest strain condition (green curve), the red and blue traces exhibit an additional peak at $0.33$~THz with amplitude within a factor of two of the TA peak at the same wavevector. The frequency of this new peak does not change appreciably between the two strained configurations (blue and red curves). Additionally, Fig.~\ref{fig5}d shows that the TO branch at $l = 0.075$~rlu softens slightly and splits into two peaks at the highest strain. This is also visible in Fig.~\ref{fig5}c, where the TO branch exhibits a splitting that is not apparent at lower strains in Figs.~\ref{fig5}a and b. 

The marked change in the phonon dispersion shown in Fig \ref{fig5}a-c is a manifestation of a drastic modification of the lattice under only moderate strain. While the soft mode associated with the traditional ferroelectric order is believed to be the TO mode at small wavevectors\cite{Cochran1960}, the most prominent change observed here occurs in the TA branch at finite wavevectors, with little change to the TO branch near the zone centre (Fig. \ref{fig4}). This is summarized schematically in Fig \ref{fig1}b-e. In a traditional ferroelectric (Fig. \ref{fig1}b) the TO mode at small wavevectors undergoes a strong renormalization (Fig. \ref{fig1}d) to produce an amplitude mode of vibration of the homogeneous polarization. The lack of softening of the TO near zone center (Fig. \ref{fig4}) and the weak softening of the TO at $l = 0.075$ rlu (Fig. \ref{fig5}d) is consistent with the anisotropy imposed by the strain, but inconsistent with a massive renormalization of the TO. On the other hand, in a textured polar state (Fig. \ref{fig1}c),  the renormalized mode of the textured (possibly fluctuating) polarization derives from an instability at finite wavevector (Fig. \ref{fig1}e)\cite{Guzman-Verri2023,Zubko2007,Muller1991}, irrespective of whether it is frozen or fluctuating. The new mode in Fig. \ref{fig5}b-c is consistent with this scenario. Because it occurs at finite wavevectors, this mode is difficult to identify from measurements that mainly sense the near-zero-wavevector response such as optical probes\cite{Li2019,Nova2019}.
Importantly, in a textured polar state the local polarization is finite but its spatial average vanishes, so no macroscopic polarization develops. The new finite-wavevector mode is nevertheless infrared active and low in frequency, and could therefore contribute to the polar response, for example to the THz-driven second-harmonic signal \cite{Li2019}, which in KTaO$_3$ has been linked to local dipolar correlations rather than global ferroelectricity~\cite{Cheng2023}, without a macroscopic polarization.
Finally, with only three strain conditions we cannot determine the order of the transition: the finite-frequency mode we observe is compatible with either a continuous or a first-order transition, and the absence of TO softening excludes only a transition driven continuously by the zone-centre TO mode.

Our results bear on a long-standing question about the nature of quantum paraelectric \sto\cite{Muller1979}: they are consistent with a scenario in which the ferroelectric instability is preempted by a competing instability at finite wavevector\cite{Muller1991,Guzman-Verri2023,Courtens1993}, whose fluctuations texture the polarization on nanometre length scales. The polar regime reached under weak uniaxial strain\cite{Uwe1976,Burke1971,Fujii1987}, is accompanied not by a collapse of the zone-centre TO mode but by a new infrared-active mode at finite wavevector, which we interpret as a possible renormalized mode of a static or fluctuating polar texture.
If this picture holds, the parent quantum paraelectric would be better understood as the fluctuating, disordered side of the transition into this textured polar state rather than a featureless centrosymmetric paraelectric held back from ferroelectric order by quantum fluctuations. Recent neutron and Raman measurements of uniaxially stressed \sto\cite{khayr2025uniaxialstraintuningpolar} likewise find the zone-centre soft TO modes underdamped and non-collapsing, consistent with our unchanged near-zone-centre TO response. That study interprets its sharp TO phonons as evidence for mesoscale disorder; our measurement locates the new mode in the acoustic sector at wavevectors $l \approx 0.06–0.13$ rlu (characteristic wavelengths 3–6.5 nm). Our approach, combining THz excitation, ultrafast x-rays, and uniaxial strain, reveals this regime through its low-energy polar collective modes, demonstrating a new route to uncovering hidden phases in quantum materials.


\section*{Methods}
\subsection*{Experimental details}

A single crystal of (001)-oriented \sto (MTI Corp) was polished and cut to $5 \times 0.2\times 0.05$ mm$^3$. 
The strain was controlled using a CS130 cell from Razorbill. The needle-shaped single crystal was suspended freely between two mounting plates of the strain cell and glued using a two-part epoxy, such as 5-minute epoxy gel and 2-ton epoxy gel from Devcon. The capacitance between the two plates provides an estimate of applied displacement.

Hard X-ray pulses with a 40 fs duration and 10.2 keV photon energy from the SwissFEL facility were monochromatized by a double crystal Silicon monochromator and focused using Kirkpatrick–Baez mirrors to a 100 $\times$ 100 $\mu m^2$ cross section at the sample.
The X-ray pulses passed through a hole in a parabolic mirror used to focus the THz pulses as shown in Fig. \ref{fig2}a.
The strain cell with the sample was mounted inside a cryogenic vacuum chamber\cite{Mankowsky_2021} with translation and rotation motions required for diffraction. The temperature was measured by a diode fixed close to the strain cell. 
Intense THz fields were generated by the pulse-front-tilt method in a LiNbO$_3$ prism\cite{Hirori2011}. The THz pulses were vertically polarized with a maximum peak electric field of around 600 kV cm$^{-1}$, as calibrated by electro-optic sampling measurements (Fig.~\ref{fig3}b inset) using a 100 $\mu$m-thick GaP crystal mounted next to the sample. Because of the high refractive index of \sto in the THz range the field inside the sample is polarized perpendicular to $[0,0,L]$. Its azimuth is $\approx 60^\circ$ to the strain axis. 
The Fourier spectrum of the THz pulse is shown in Fig.~\ref{fig:s1}. The THz beam was focused to an $800 \times 800$ $\mu m^2$ beam size on the sample collinearly with the X-ray beam using a parabolic mirror. 

X-ray diffraction patterns around $(3,3,3)$ diffraction peak were acquired as a function of the delay between THz pump and X-ray probe pulses. The scattered photons were recorded by a Jungfrau area detector positioned at 180 mm from the sample outside of the cryogenic chamber. To match the pumped and probed volumes while maximizing the pump fluence, we implemented a grazing exit geometry. The exit angle was set to about 1$^{\circ}$ with respect to the sample surface such that the detected X-ray photons originated from a depth of  $<1 \mu$m from the sample surface, comparable to the penetration depth of the THz pulse in \sto at low temperatures. The sample was rotated around an axis perpendicular to the incident beam to achieve the asymmetric Bragg condition for $(3,3,3)$, and then around the incident beam to reach the desired exit angle. The shot-by-shot detector images were grouped into laser-on and laser-off data sets and binned into 50 fs time bins. Typically, around 1300 events were accumulated for each data point. 

The data in Figs.~\ref{fig5}a-c were recorded at a fixed sample angle slightly detuned from the (3,3,3) Bragg condition. Each detector pixel samples a different momentum transfer on the Ewald sphere, which intersects the truncation rod near $l\approx0.1$ and passes within $2.3\times10^{-3}$~rlu of it for $0.04\le l\le0.14$, where the rod intensity remains above $\sim$15\% of its on-rod value (\ref{fig:s7}). The data in Fig.~\ref{fig4} were recorded at the Bragg condition, where the sphere passes through the rod at $l=0$ and stays within $1.6\times10^{-3}$~rlu up to $l=0.06$. 
In this geometry the detector saturates for $l \leq 0.01$ rlu because of the strong Bragg peak; the Bragg-peak response in Fig. 3c (dark blue line) was measured separately.

\subsection*{Strain details}
The Razorbill CS130 strain cell uses parallel plate capacitive sensors to measure the displacement of the sample. The capacitance-displacement relation was calibrated at room temperature, and the capacitance at zero tension was calibrated as a function of temperature. The effective length $L$ is defined as the distance between two extremes of the sample glued to the cell plates. The displacement $\Delta L$ is obtained from the capacitance reading. The relative displacement $\Delta L/L$ is the control parameter and is used to label the strain condition in the main figures. 
To estimate the strain on the illuminated spot of the sample we compute the average change in the $(3,3,3)$ lattice plane spacing $\Delta d_{(3,3,3)}$ from the measured scattering angle (``two-theta'') and Bragg's law. The anisotropic strain $\epsilon$ can be estimated from $\Delta d_{(3,3,3)}$ as follows: Taking Poisson ratio $\sigma$\cite{scott1997interpretation} and assuming for simplicity a cubic unit cell that changes under strain as $a \to a(1+\epsilon)$, $b \to a (1-\sigma \epsilon)$ and $c \to a (1-\sigma \epsilon)$ we obtain 
\[\Delta d/d = [(h^2 - (k^2 + l^2) \sigma)/(h^2 + k^2 + l^2)] \epsilon + O(\epsilon^2).
\]
This results in $\Delta d(3,3,3)/d(3,3,3) = 1/3 (1 - 2 \sigma) \epsilon + O(\epsilon^2)$, which for $\sigma = 0.2$ gives $\epsilon \approx 5 \Delta d/d$ as plotted in Fig. \ref{fig1}a. We estimated the error on each shift of the Bragg peak on the detector by the standard error of the fitted peak centre, which gives $\epsilon=(1.30\pm0.01)\times10^{-3}$ and $(1.45\pm0.01)\times10^{-3}$. For comparison, $\sigma = 0$ yields $\epsilon \approx 3 \Delta d/d$. In either case the two nonzero strain markers remain inside the polar region of Fig. \ref{fig1}a.
In the main text we report $\Delta d_{(3,3,3)}/\bar{d}_{(3,3,3)}$, where $\bar{d}_{(3,3,3)}$ is the nominal $(3,3,3)$ lattice plane spacing of the unstrained sample. The reported values $\Delta d_{(3,3,3)}/\bar{d}_{(3,3,3)}$ do not depend on the exact value of $\bar{d}_{(3,3,3)}$ or the Poisson ratio.

\section*{Data availability}
The data supporting the findings of this study are reported in the main text, Methods and Extended Data. Raw data are available from the corresponding author upon reasonable request.


\section*{Acknowledgements}
We acknowledge the Paul Scherrer Institute, Villigen, Switzerland for provision of free-electron laser beamtime at the Bernina instrument of the SwissFEL ARAMIS branch.
H.W., J.S., D.A.R. and M.T. were supported by the US Department of Energy, Office of Science, Office of Basic Energy Sciences through the Division of Materials Sciences and Engineering under Contract No. DE-AC02-76SF00515. E.F. was supported by N.S.F. fellowship DGE-2146755. P.M. was supported by an XtalPi AI for Science fellowship. G.O. acknowledges support from the Koret Foundation.


\section*{Author contributions}
M.T. conceived and supervised the project. H.W. led the experiment. 
H.W., J.S., E.F., G.O., P.M., L.F., R.M., S.W.H., M.S., H.L., S.Z., M.T. performed the experiment.
M.M. and H.W. prepared the samples. H.W., G.O. and M.T. wrote the initial manuscript with feedback from all authors.


\section*{Competing interests}
The authors declare no competing interests.


\bibliographystyle{naturemag}
\bibliography{strainedSTOmain}


\clearpage
\setcounter{figure}{0}
\renewcommand{\thefigure}{Extended Data \arabic{figure}}

\begin{figure}
	\centering
	\includegraphics[width=0.6\textwidth]{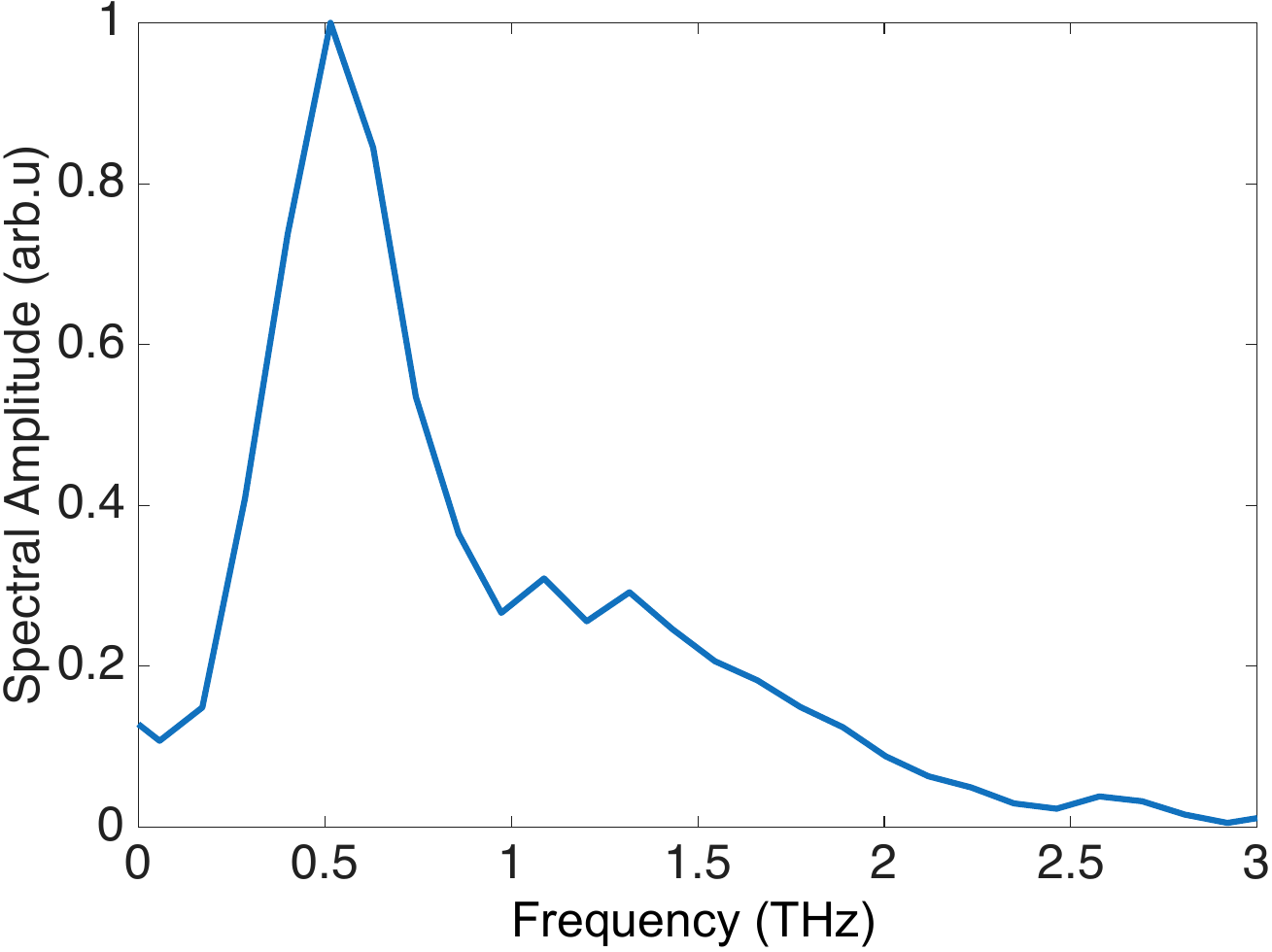} 
	\caption{\textbf{THz pump field spectrum.}
		Fourier transform of the electro-optic sampling trace shown in the inset of Fig.~\ref{fig3}b.}
	\label{fig:s1} 
\end{figure}

\begin{figure}
	\centering
	\includegraphics[width=1\textwidth]{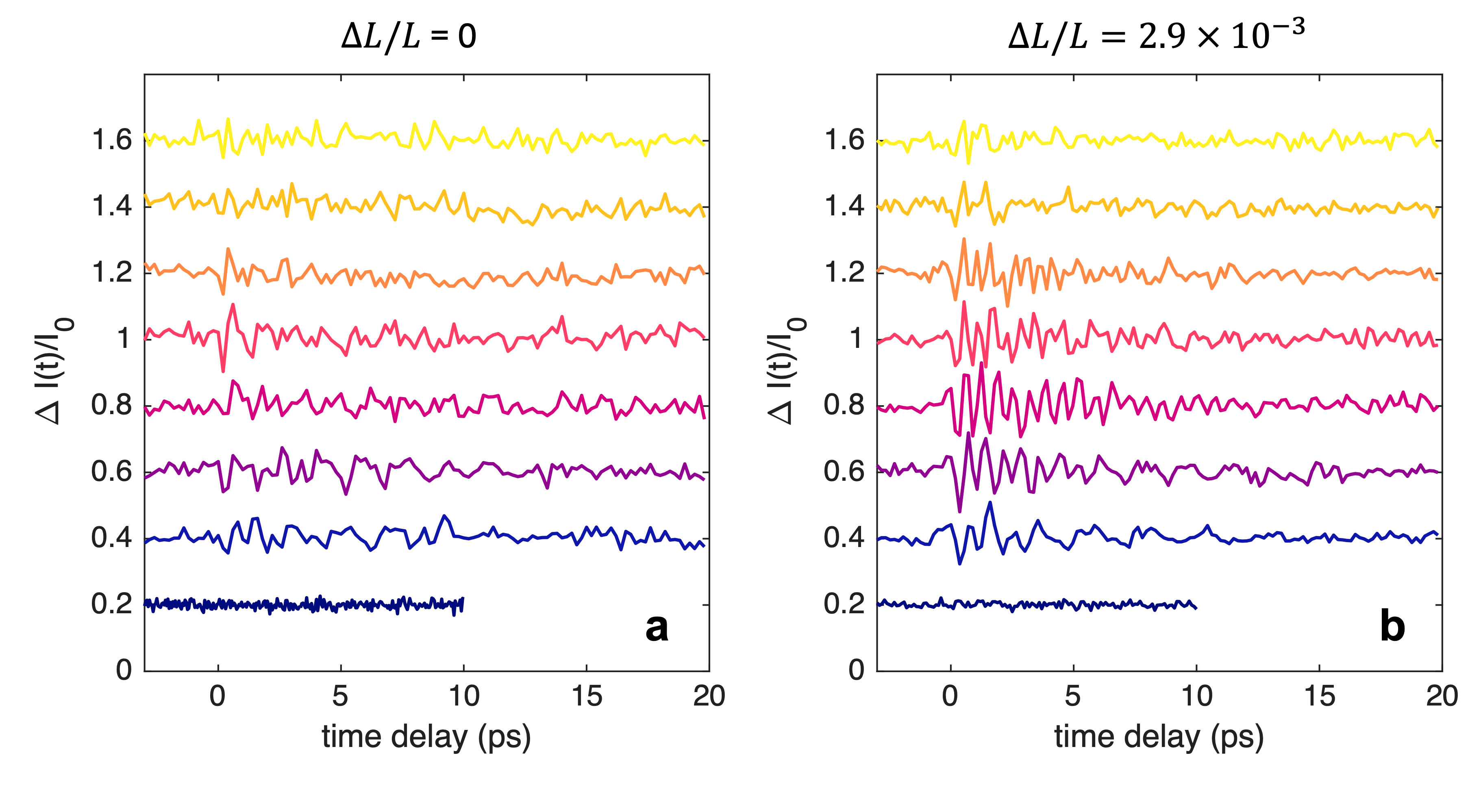} 
	\caption{\textbf{Relative diffraction intensity changes as a function of wave vector, delay time and strain.}
		Relative change in diffraction intensity ($\Delta I(t)/I_0$) for representative wavevectors along the [0,0, $l$] direction indicated by the color-coded dots in Fig.~\ref{fig3}b at (\textbf{a}) $\Delta L/L=0$ and (\textbf{b}) $\Delta L/L=2.9 \times 10^{-3}$ strain conditions.}
	\label{fig:s2}
\end{figure}

\begin{figure} 
	\centering
	\includegraphics[width=0.9\textwidth]{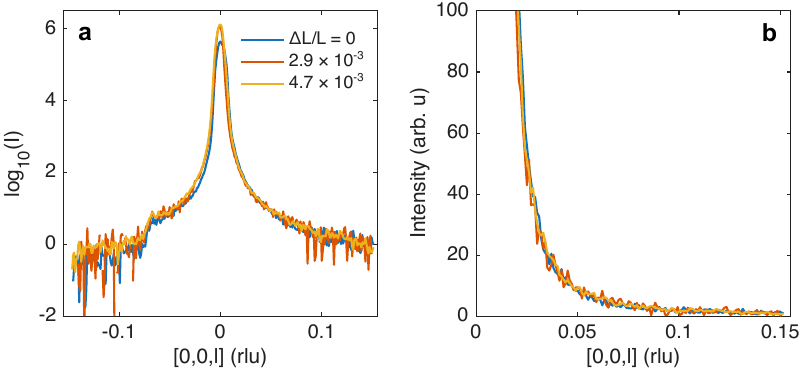} 
	\caption{\textbf{Intensity along (00L) for the three strain conditions.}
		\textbf{a}-\textbf{b} show the intensity of the (00L) cut of the (3,3,3) Bragg peak for several strain configurations labeled by $\Delta L/L$ measured at the Swiss FEL. \textbf{a} and \textbf{b} show the logarithmic (base 10) and linear scale, respectively.}
	\label{fig:s3} 
\end{figure}

\begin{figure} 
	\centering
	\includegraphics[width=0.99\textwidth]{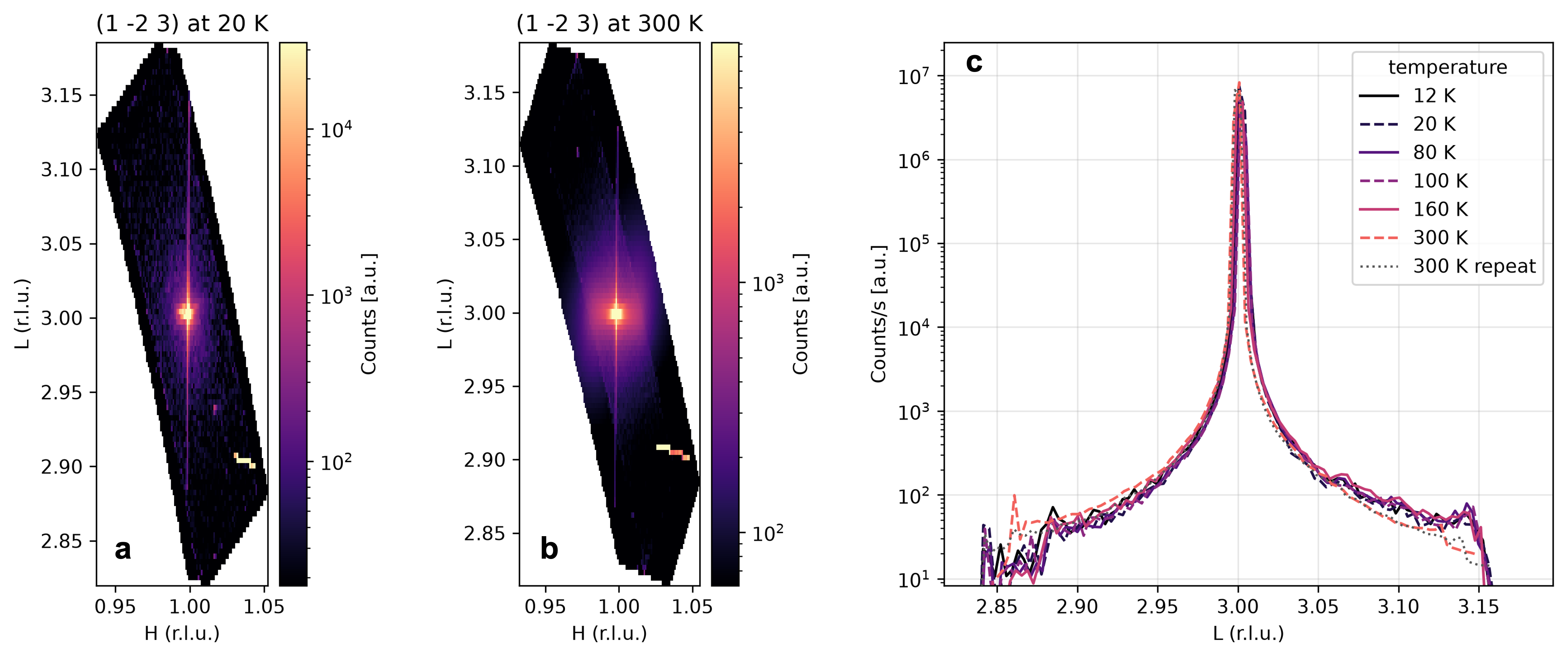} 
	\caption{\textbf{Temperature dependence of scattering streak.} Complementary measurements at the Stanford Synchrotron Radiation Lightsource (SSRL) in the same grazing-exit geometry as the SwissFEL experiment. (\textbf{a-b}) Reciprocal-space map around the (1, -2, 3) Bragg peak at 20 K and 300 K, respectively, showing the same elongated streak along [0, 0, l] observed at (3, 3, 3) (Fig. 3a). Its appearance at multiple Bragg reflections (not shown) is characteristic of a crystal truncation rod from the sample surface. (\textbf{c}) Intensity of the (00L) cut at representative temperatures. The intensities were scaled to account for slight variations of the incident flux and sample position variations.
    The intensity profiles are virtually independent of temperature, and likewise independent of strain (Extended Data 3), as expected for a surface truncation rod.
        }
	\label{fig:s4} 
\end{figure}

\begin{figure}
	\centering
	\includegraphics[width=0.9\textwidth]{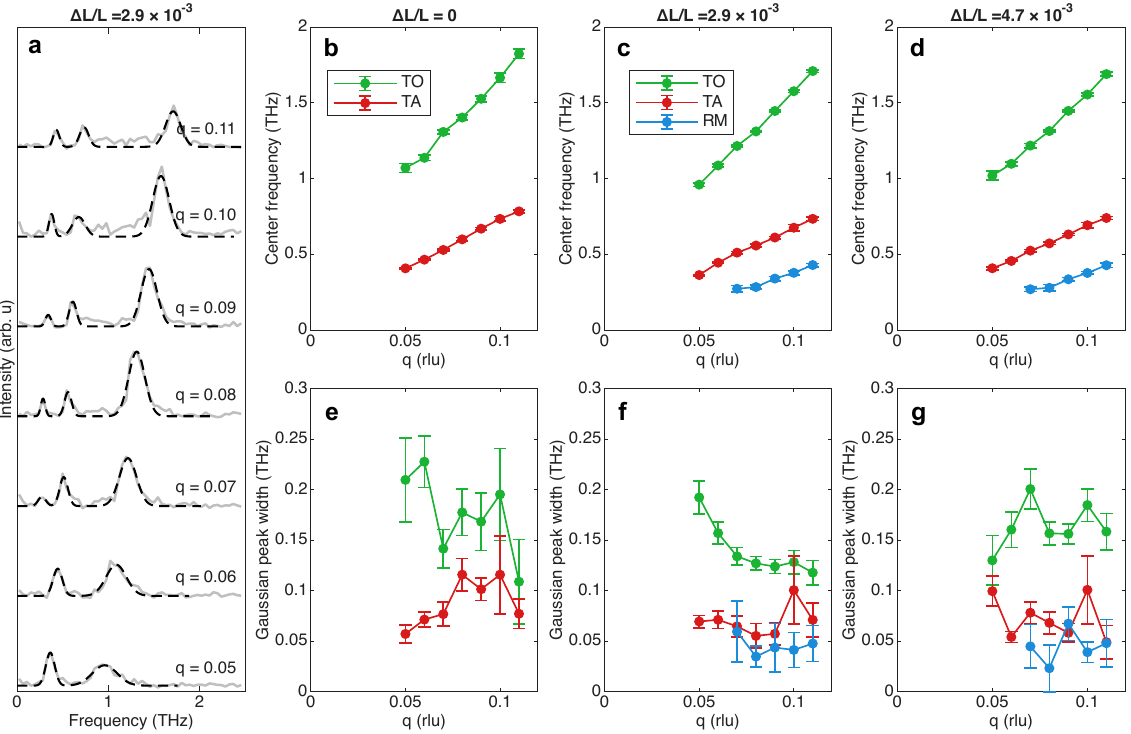} 
	\caption{\textbf{Gaussian fits to the dispersion.} \textbf{(a)} spectrum at representative wavevectors for $\Delta L/L = 2.9\times 10^{-3}$ (gray solid lines). Dashed lines show fit with three Gaussians. Panels \textbf{(b-d)} show the centre of the gaussian fit for the three strain conditions, respectively, and \textbf{(e-g)} show the corresponding standard deviation. The error bars are the standard errors of the fit.
        }
	\label{fig:s5} 
\end{figure}

\begin{figure}
\centering
\includegraphics[width=0.5\textwidth]{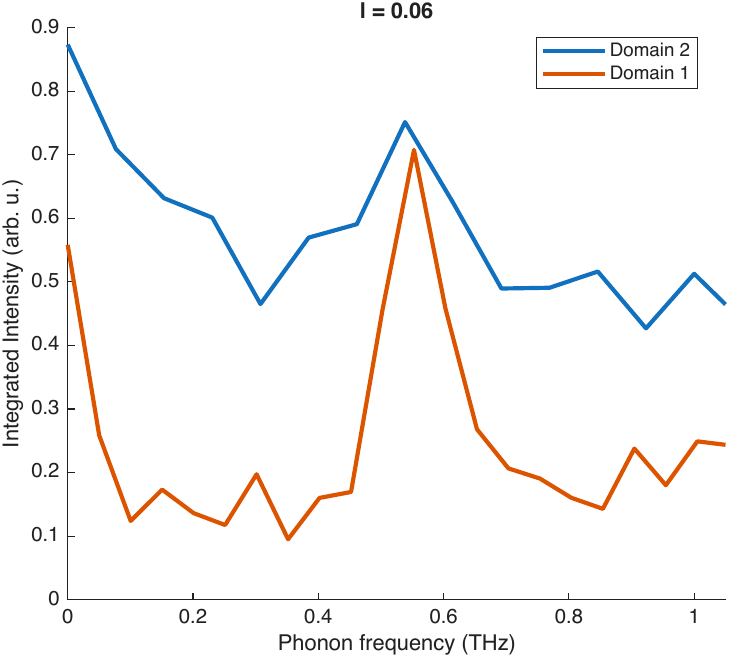}
\caption{\textbf{Phonon spectra of the two tetragonal domains in the unstrained sample.}
Spectra at $l = 0.06$~rlu for the two domain orientations, which appear
as separate peaks in the rocking scan and are isolated on the detector (Fig.~\ref{fig1}b). Both domains shows a single TA peak near 0.55 THz and no mode at the frequency of the strain-induced branch. This shows that the new branch that appears in Fig. \ref{fig5}b-c is not due to domain averaging.}
\label{fig:s6}
\end{figure}

\begin{figure}
\centering
\includegraphics[width=0.65\textwidth]{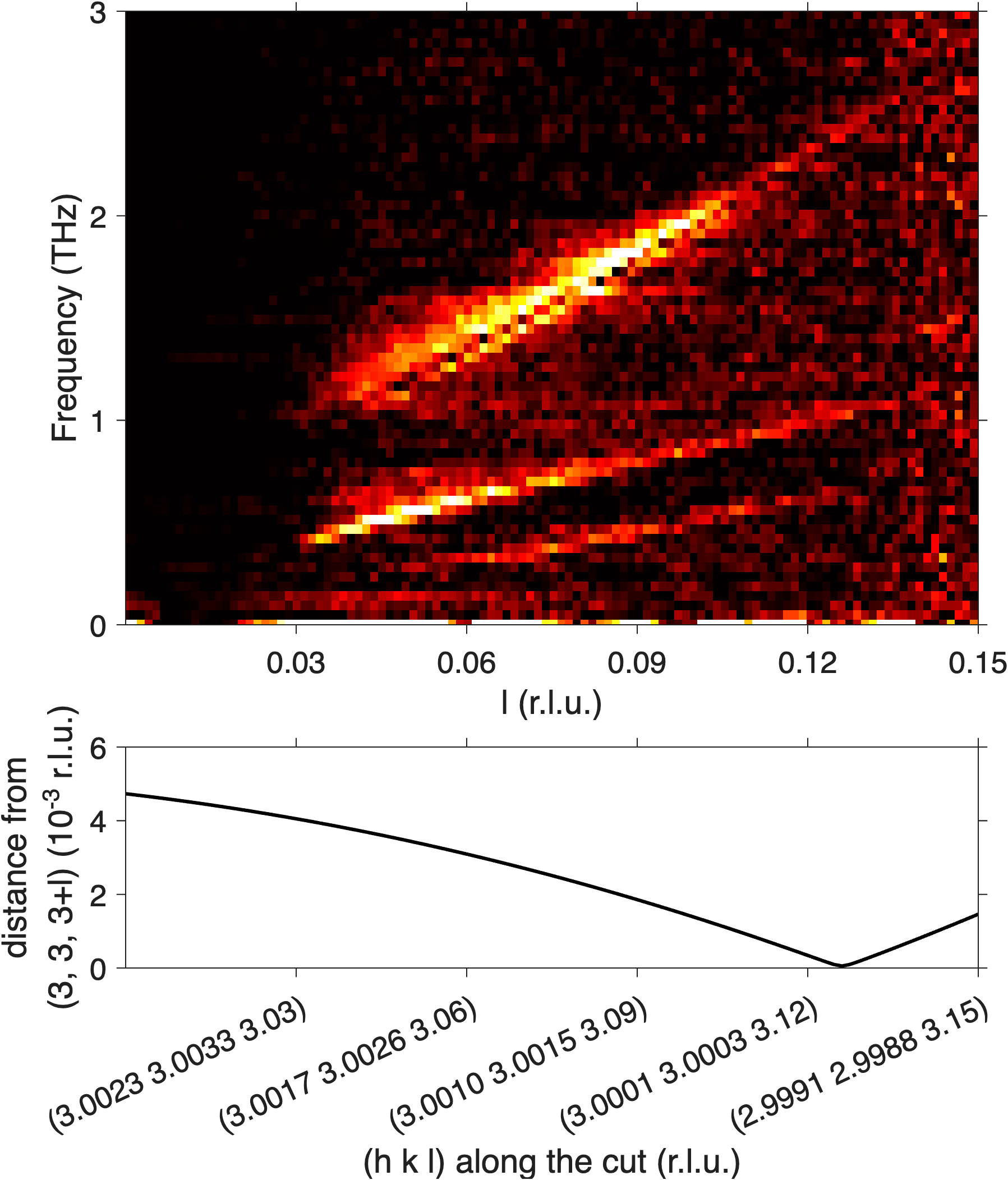}
\caption{\textbf{Momentum transfer probed in the detuned geometry relative to the
truncation rod.} Phonon spectrum in the same geometry and strain condition as Fig.~\ref{fig5}c (top panel). In-plane distance of the probed $\mathbf Q$ from the truncation rod $(3,3,3+l)$ along the cut (bottom panel). 
}
\label{fig:s7}
\end{figure}

\end{document}